\documentclass[twocolumn,preprintnumbers,superscriptaddress,nofootinbib,aps,prd,floatfix]{revtex4}

\usepackage{footmisc,multirow}
\usepackage{subfigure,color}
\usepackage{amsmath,slashed,enumerate}

\usepackage{graphicx}

\newcommand{\be}{\begin{eqnarray*}}
\newcommand{\ee}{\end{eqnarray*}}
\newcommand{\gl}[1]{(\ref{#1})}

\newcommand{\bee}{\begin{eqnarray}}
\newcommand{\eee}{\end{eqnarray}}
\newcommand{\beeq}{\begin{equation}}
\newcommand{\eeeq}{\end{equation}}

\newcommand{\ep}{\varepsilon}
\renewcommand{\vec}{\bf}

\preprint{IPPP/14/11} \preprint{DCPT/14/22} \preprint{IPMU15-0039}

\begin{document}

\title{Unitarity-controlled diboson resonances after Higgs discovery}

\begin{abstract}
  If the recently discovered Higgs boson's couplings deviate from the
  Standard Model expectation, we may anticipate new resonant physics
  in the weak boson fusion channels resulting from high scale
  unitarity sum rules of longitudinal gauge boson
  scattering. Motivated by excesses in analyses of
  multi-leptons+missing energy+jets final states during run 1, we
  perform a phenomenological investigation of these channels at the
  LHC bounded by current Higgs coupling constraints. Such an approach
  constrains the prospects to observe such new physics at the LHC as a
  function of very few and generic parameters and allows the
  investigation of the strong requirement of probability conservation
  in the electroweak sector to high energies. Our analysis is directly
  relevant for the 2 TeV excess reported recently by the CMS and ATLAS
  collaborations.
\end{abstract}

\author{Christoph Englert} \email{christoph.englert@glasgow.ac.uk}
\affiliation{SUPA, School of Physics and Astronomy, University of
  Glasgow,\\Glasgow, G12 8QQ, United Kingdom\\[0.1cm]}

\author{Philip Harris} \email{philip.coleman.harris@cern.ch}
\affiliation{CERN, CH-1211 Geneva 23, Switzerland\\[0.1cm]}

\author{Michael Spannowsky} \email{michael.spannowsky@durham.ac.uk}
\affiliation{Institute for Particle Physics Phenomenology, Department
  of Physics,\\Durham University, Durham, DH1 3LE, United Kingdom\\[0.1cm]}

\author{Michihisa Takeuchi} \email{michihisa.takeuchi@kcl.ac.uk}
\affiliation{Kavli IPMU (WPI), The University of Tokyo, Kashiwa,
  277-8583, Japan\\[0.1cm]}

\maketitle


\section{Introduction}
\label{sec:intro}

After the discovery of the Higgs boson~\cite{orig} at the LHC and
first preliminary tests of its coupling structure and
strengths~\cite{hatlas,hcms}, a coarse-grained picture of consistency
with the Standard Model (SM) has emerged. Resulting from Higgs quantum
numbers, current constraints on the Higgs boson's couplings, assuming
a SM-value of the Higgs width or an upper limit on the Higgs' coupling
to electroweak gauge bosons indicate that the Higgs couplings to
electroweak bosons agree with the SM expectation within
${\cal{O}}(10\%)$~\cite{sven,sfitter}. This establishes the Higgs'
involvement in electroweak symmetry breaking and its role in the
unitarization of massive longitudinal gauge boson scattering.

However, current constraints leave a lot of space for deviations from
the SM-like implementation of electroweak symmetry breaking. In
particular, small deviations from the SM Higgs coupling pattern are
expected in a very broad class of models that explain the presence of
the electroweak scale as a dimensional transmutation
effect~\cite{Giudice:2007fh,conformal}. In particular, these include
composite Higgs scenarios where we expect new contributions from
composite states analogous to the rho meson
\cite{Bellazzini:2012tv}. Explicit examples have been discussed in the
literature, mostly in the context of AdS/CFT duality, see {\it
  e.g.}~\cite{gaugephobic}.

Owing to the fact that any modification from the SM Higgs couplings
explicitly introduces unitarity violation, novel resonant physics is
likely to enter at a scale $Q^2\gg m_h^2$ to conserve
probability~\cite{Cornwall:1974km} if we indeed deal with non-SM Higgs
interactions. Weak boson scattering processes are theoretically
well-motivated probes of such dynamics, correlating the size of the
new physics effects with the deviation of the observed Higgs
phenomenology from the SM.

Accessing longitudinal gauge boson scattering (which is highly
sensitive to BSM effects) at the LHC in a phenomenologically useful
way is difficult. Due to almost conserved light quark and lepton
currents, weak boson fusion (WBF, for analyses
see~\cite{wbf_old,wbf_new}) is not too sensitive to modifications of
the involved Higgs couplings.\footnote{In a general gauge the
  Goldstone contributions to the amplitude vanish in the chiral limit,
  signalling a vanishing contribution from longitudinal degrees of
  freedom at high invariant masses.} The Higgs exchange at energies
$m(VV)\gg m_h$ in a Higgs doublet model provides a destructive
contribution to $VVqq$ ($V=W^\pm,Z$) production. Thus, a $\sim 10\%$
cross section excess at the LHC for inclusive WBF is mainly due to the
smaller destructive Higgs contribution for smaller couplings, rather
than diverging $qq\to qq VV$ processes getting tamed by the polynomial
parton density function suppression at large parton energy fractions.

\begin{figure*}[!t]
  \begin{center}
    \includegraphics[width=0.65\textwidth]{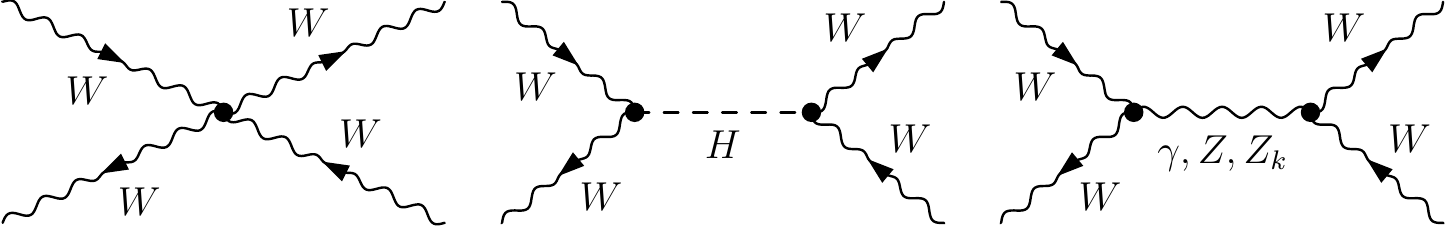}
    \caption{\label{fig:feyn} Sample Feynman diagrams contributing to
      $WW\to WW$, the $t$-channel diagrams are not shown.}
  \end{center}
\end{figure*}

Nevertheless, it is important to realize that, if $V_LV_L\to V_LV_L$
(Fig.~\ref{fig:feyn}) scattering violates the unitarity bound, the
(leading order) electroweak sector becomes ill-defined, and there is
no theoretically consistent interpretation of constraints and
measurements even if the alternate hypothesis seems
well-behaved~\cite{Englert:2014aca}.

Current analyses mostly focus on studying the impact of a subset of
the 59 dimension-six operators (neglecting flavor
structures)~\cite{dim6} on Higgs physics in the on- and off-shell
region. In this paper, we take a complementary approach and address
the question of what to expect in WBF processes when unitarity is
explicitly enforced by additional resonances in the TeV regime,
following a strong-interaction paradigm.

If additional resonances in $VV$ scattering are present, an
identification will depend on their mass, width and coupling
strengths, fixed through high scale unitarity as a function of their
spin: The naive growth proportional to $s^2$ and $s$ of the amplitude,
depicted in Fig.~\ref{fig:feyn}, in the high energy limit
$\ep_L^\mu(p)\sim p^\mu/m_V$ is mitigated by imposing sum rules that
link quartic and trilinear gauge and Higgs couplings (see
also~\cite{perelstein_matchev,csaki,Bhattacharyya:2012tj} for a
similar discussion of the pure Higgs-less case). 

For SM-like $WW$ scattering, the sum rules read
\begin{subequations}
  \label{eq:unitr}
  \begin{widetext}
    \begin{alignat}{5}
      \label{ww1}
      g_{WWWW}&= g_{WW\gamma}^2+\sum_i g_{WWZ_i}^2 \\
      \label{ww2}
      4m_W^2g_{WWWW} &= \sum_i 3 m_i^2 g_{WWZ_i}^2
      + \sum_i g_{WWH_i}^2 \,,
    \end{alignat}
    and for $WW\to ZZ$ (and crossed) scattering these are modified to
    \begin{alignat}{5}
      \label{wz1}
      g_{WWZZ}& =\sum_i g_{W_iWZ}^2  \\
      \label{wz2}
      2(m_W^2+m_Z^2)g_{WWZZ} &= \sum_i \left(3 m_i^2 - {(m_Z^2-m_W^2)^2\over
          m_i^2} \right) g_{W_iWZ}^2 + \sum_i g_{WWH_i}g_{ZZH_i}  \,.
    \end{alignat}
  \end{widetext}
\end{subequations}
In these sums the index $i=1$ refers to the SM $W$, $Z$ and Higgs
bosons, respectively, and $i>1$ refer to a series of isotriplet
massive vector bosons $W'$, $Z'$ and isosinglet $H'$ scalar bosons
respectively.\footnote{It is worth noting that similar sum rules
  cannot be formulated for iso-tensors~\cite{juergen}.}  Although we
will not make contact with a concrete model, one can think of the
$i>1$ states as Kaluza-Klein states that arise in models with extra
dimensions and dual interpretations thereof~\cite{csaki,gaugephobic}
as a guideline: $W_{i>1}$ can couple to SM $W$ and $Z$ bosons, while
$Z_{i>1}$ can couple to a pair of SM $W$ bosons etc. In concrete
scenarios \cite{csaki,gaugephobic,Bellazzini:2012tv} the above sum
rules are quickly saturated by the first $i\neq 1$ states. We assume
that custodial $SU(2)$ is intact, which, in addition to the correct
tree-level $Z/W$ mass ratio, will leave imprints in the the additional
resonance’s spectrum, see e.g. \cite{gaugephobic}. The unitarity sum
rules are independent of custodial isospin and since the sum rules are
quickly saturated, custodial $SU(2)$ is not important for our
investigation, but remains a testable concept in case of a discovery
of additional vector resonances.

The discovery of particles categorized as Eq.~\eqref{eq:unitr} in the
$VVjj$ channels would provide a conclusive hint for the role of new
resonances in electroweak symmetry breaking. It is intriguing that
both ATLAS and CMS have observed non-significant excesses in
(multi-)lepton+$\slashed{E}_T$+jets searches~\cite{exibump}.

In addition, recently, both ATLAS and CMS reported on excesses in
final states with reconstructed hadronically-decaying di-vector boson
final states with an invariant mass $m_{VV} \simeq 2$ TeV
\cite{atlasex,cmsex}. ATLAS found a global significance of $2.5$
standard deviations. Both vector bosons were reconstructed using fat
jets and jet substructure methods, i.e. mass-drop and filtering
\cite{Butterworth:2008iy}. While WBF tagging jets are very energetic,
they have small transverse momentum. Hence, they are likely to be
overlooked in the reconstruction procedure applied.  We take this
observation as another motivation for an as model-independent as
possible analysis of these final states.

It is important to realize that due to $SU(2)_L$ invariance ({\it
  e.g.} the absence of a quartic $Z$ interaction) the reasoning along
the above lines does not apply to $ZZ\to ZZ$ scattering. In the high
energy regime the Higgs exchange diagrams conspire
\begin{equation}
  {\cal{M}}(Z_LZ_L\to Z_LZ_L) \sim s+t+u = 4m_Z^2\,,
\end{equation}
{\it i.e.} the scattering amplitude becomes independent of the center
of mass energy. Hence, on the one hand, in scenarios where unitarity in $WW$
and $WZ$ scattering is enforced by iso-vectors, we do not expect new
resonant structures in $pp\to 4\ell +2 j$. On the other hand if unitarity is conserved
via the exchange of iso-scalar states, this channel will provide a
phenomenological smoking gun. Obviously this is not a novel insight
and under discussion in the context of {\it e.g.} Higgs portal
scenarios~\cite{portal}. We will not investigate the $ZZ$ channel
along this line in further detail.

For the purpose of this paper we start with a minimal, yet powerful
set of assumptions, that can be reconciled in models that range from
(perturbative and large $N$) AdS/CFT duality over SUSY to simple Higgs
portal scenarios. We will focus on a vectorial realization of
unitarity, assuming an electroweak doublet nature of the Higgs
boson.\footnote{See~\cite{trip} for a detailed discussion of WBF
  signatures in Higgs triplet scenarios.} This represents an
alternative benchmark of new resonant physics involved in the
mechanism of EWSB which has been largely ignored after the Higgs
discovery so far.

The first rules Eq.~\gl{ww1},~\gl{wz1} are typically a consequence of
gauge invariance~\cite{csaki} while the second rules
\gl{ww2},~\gl{wz2} reflect the particular mechanism of EWSB. Similar
sum rules exist for massive $q\bar q \to V_LV_L$ scattering, linking
the Yukawa sector to the gauge sector~\cite{chanowitz}. We are
predominantly interested in a modified Higgs phenomenology in the
standard WBF search channels. It is however important to note that the
latter sum rules also predict new resonant states in Drell-Yan type
production~\cite{drell} (for a recent comprehensive discussion see
also~\cite{ricardo}) or gluon fusion induced $VVjj$ production. For
this analysis, gluon fusion events can efficiently be removed by
imposing selection criteria~\cite{dieter}; this process is neglected
further on (see below).

\begin{figure}[!t]
  \begin{center}
        \includegraphics[width=0.47\textwidth]{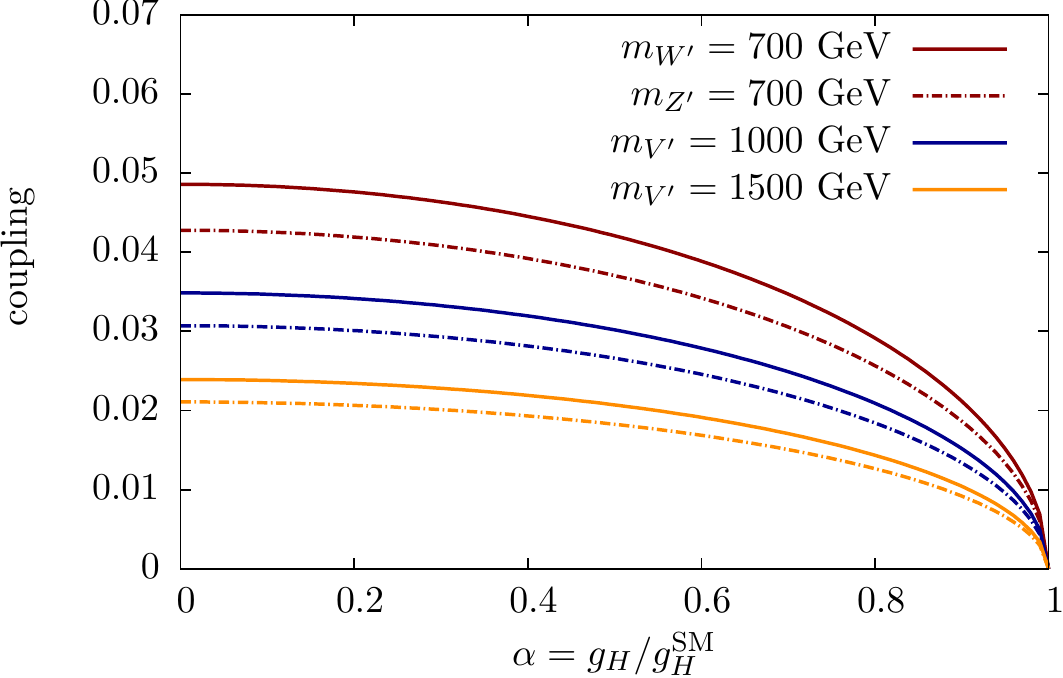}
        \caption{\label{fig:coupling}$W'$ and $Z'$ couplings to SM $W$
          and $Z$ bosons as function of the Higgs coupling deviation
          following from Eq.~\eqref{eq:unitr}.}
  \end{center}
\end{figure}

The presence of unitarizing spin one resonances is tantamount to a
modification of the 4-point gauge interactions when we choose the
trilinear couplings to be SM-like. In higher dimensional and dual
composite Higgs scenarios this fact is typically encoded in multiple
definitions of the tree-level Weinberg angle and a resulting
constraint from the $\rho$ parameter. The quartic gauge couplings are
currently not well constrained and we use this freedom to saturate the
above sum rules via a non-standard value of $g_{WWWW}$ and
$g_{WWZZ}$. The numerical modifications away from the SM values as a
function of the modified Higgs couplings is small $\simeq 0.1\%$,
especially in the vicinity of the SM when $g_{WWZ'}=g_{W'WZ}=0$ are
small and well within the latest quartic coupling measurements'
uncertainty as performed during the LEP
era~\cite{quartic}.\footnote{On a theoretical level, a modification of
  the quartic interactions away from the the SM expectation introduces
  issues with Ward identities which ultimately feed into the unitarity
  of the $S$ matrix beyond the tree-level approximation. Hence,
  Eqs.~\gl{eq:unitr} need to be understood as an effective theory
  below the compositeness scale. In concrete scenarios motivated from
  AdS/CFT, the fundamental scale can be as high as 10~TeV
  \cite{csaki,gaugephobic} and the SM-like ward identities need to be
  replaced by the corresponding 5d AdS relations.}

\section{Results}
\label{sec:results}

\subsection{Details of the simulation}
\label{sec:sim}
Using Eq.~\eqref{eq:unitr}, we have a simple parameterization of new
physics interactions in terms of mass and width of the new vector
state, and Higgs coupling modification parameter. Since we do not
specify a complete model we treat the extra boson widths as nuisance
parameters. In concrete models the width can span a range from rather
narrow to extremely wide. Masses are typically constrained by
electroweak precision measurements. Since the sum rules give an
independent prediction, we will not consider these corrections further.

We use a modified version of {\sc{Vbfnlo}}~\cite{vbfnlo} to simulate
the weak boson fusion channel events for fully partonic final states
inputting the relevant model parameters mentioned above.  Since WBF
can be identified as ``double-DIS'' we can efficiently include the
impact of higher order QCD corrections on differential distributions
by dynamically choosing the $t$-channel momentum transfer of the
electroweak bosons as the factorization and renormalization
scales~\cite{wbfscales} irrespective of new resonant structures in the
leptonic final state~\cite{Englert:2008wp}. We generate the gluon
fusion contribution using again {\sc{Vbfnlo}}, but find that they are
negligible for typical WBF requirements. As benchmarks we consider the
following parameter points, defining $\alpha=g_H/g_H^{\text{SM}}$,
\begin{alignat}{5}
m_{W',Z'}=700~\text{GeV}, &\quad& \Gamma_{W',Z'}=3~\text{GeV}, &\quad& \alpha=0.9, \notag\\
m_{W',Z'}=1000~\text{GeV}, &\quad& \Gamma_{W',Z'}=7~\text{GeV}, &\quad& \alpha=0.9, \notag\\
m_{W',Z'}=700~\text{GeV},  &\quad& \Gamma_{W',Z'}=10~\text{GeV}, &\quad& \alpha=0.5,\notag \\
m_{W',Z'}=1000~\text{GeV}, &\quad & \Gamma_{W',Z'}=30~\text{GeV}, &\quad& \alpha=0.5.
\end{alignat}
to highlight characteristics. Note that the values $\alpha>1$ do not
allow real coupling values in Eq.~\eqref{eq:unitr} and we cannot
incorporate this situation in model where the Higgs boson is a
$SU(2)_L$ doublet. The chosen width values are small and neglect
potentially large couplings to fermions, especially to the top
quark. We use these values to establish an estimate on sensitivity for
a particular resolution around the vector boson candidate's mass in
Sec.~\ref{sec:limits}. As we will see in Sec.~\ref{sec:limits}, where
we generalise away from the above assumptions, the signal quickly
decouples.\footnote{It is important to stress that new sources of
  theoretical uncertainties arise once the width becomes comparable to
  the resonance mass~\cite{Stuart:1991xk}.}

The {\sc{Vbfnlo}} event files are further processed with
{\sc{Herwig++}}~\cite{Bahr:2008pv} for showering and
hadronization. For this study, we utilize leptonic final states
exclusively at 14 TeV. As potential backgrounds we consider continuum
$WW$, $WZ$ and $t\bar t$ production and generate these events using
{\sc{Alpgen}}~\cite{alpgen}.

Detector effects and reconstruction efficiencies are performed using a
detector simulation based on the ATLAS Krakow
parameterization~\cite{detector}. The parameters employed provide
conservative estimates of the ATLAS detector performance for the
phase-II high-luminosity LHC. In particular we model pile-up (at $\mu
= 80$) and $\sum E_T$ dependent resolutions for jets and for $p_T$.

Jets are reconstructed with the anti-$k_T$ jet clustering
algorithm~\cite{Cacciari:2008gp} with $p_T>40$ GeV and resolution
parameter $R=0.4$. To parameterize jet resolutions, b-jet efficiencies
and fake rates we follow~\cite{detector} as well.

Charged leptons (electrons and muons) are considered to be isolated if
$p_{T,l} > 10$ GeV and if the hadronic energy deposit within a cone of
size $R = 0.3$ is smaller than $10\%$ of the lepton candidate's
transverse momentum in the rapidity range $|y_l| < 2.5$.

\begin{figure}[!t]
  \begin{center}
    \subfigure[\label{fig:ww} Transverse mass distribution of the
    $2l+\slashed{E}_T+2j$ final state after requesting exactly two isolated leptons, as outlined in Sec.~\ref{sec:2l}.]{\includegraphics[width=0.47\textwidth]{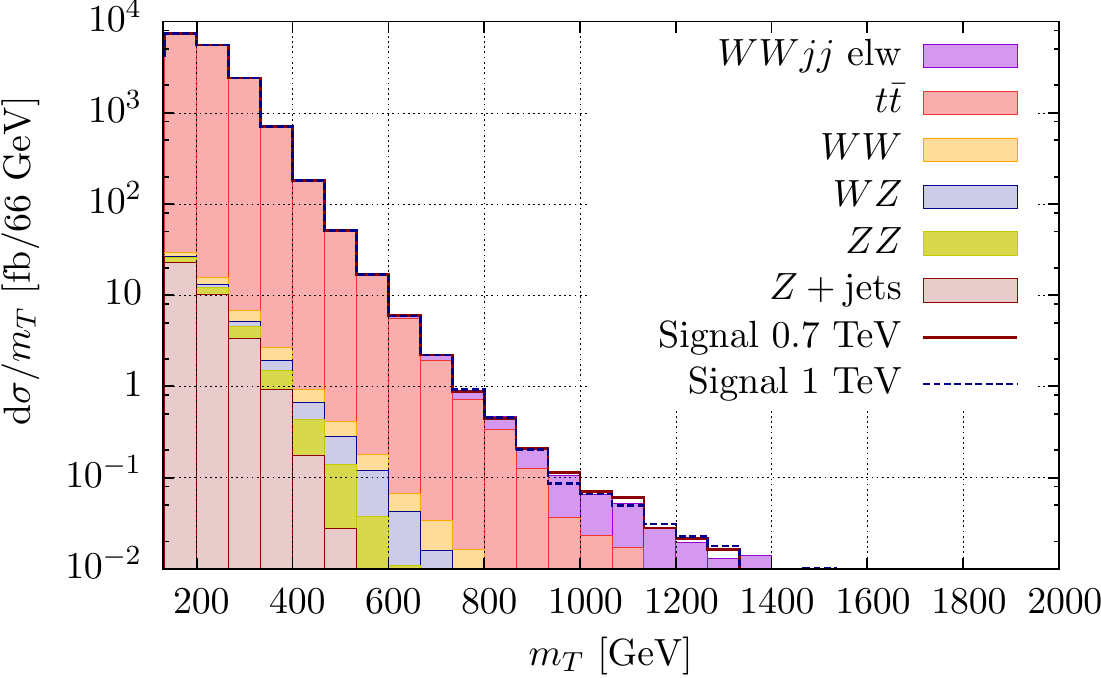}}\\
    \subfigure[\label{fig:wz} Transverse mass distribution of the
    $3l+\slashed{E}_T+2j$ final state after requesting exactly three isolated leptons, as outlined in Sec.~\ref{sec:2l}.]{\includegraphics[width=0.47\textwidth]{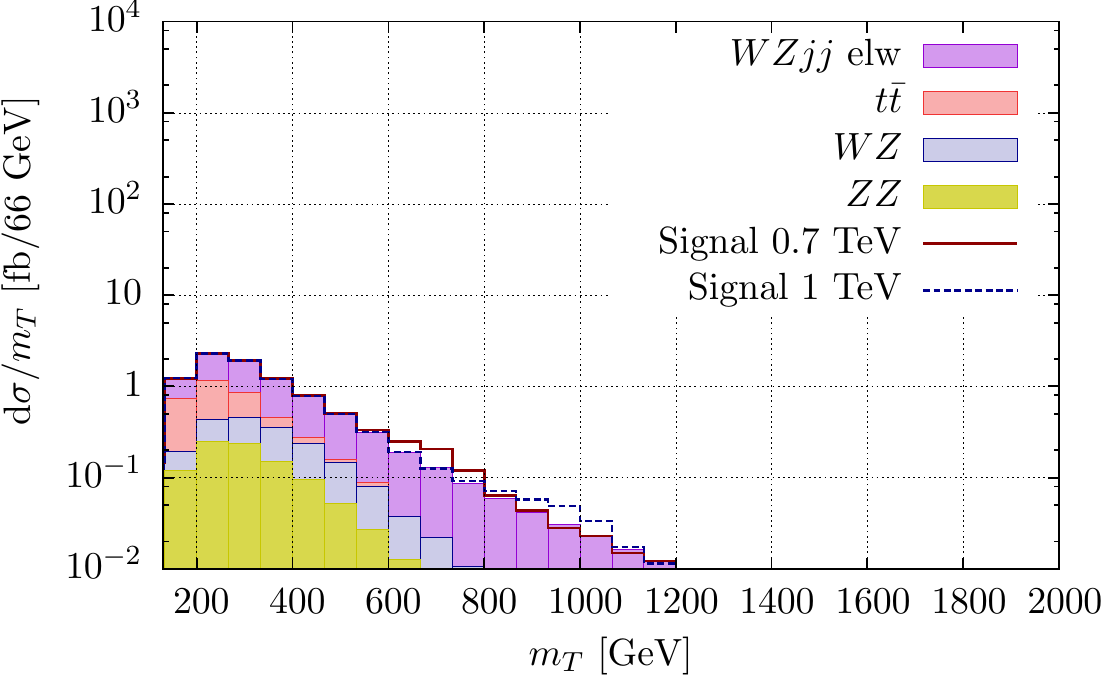}}\\
    \subfigure[\label{fig:zz} Transverse mass distribution of the
    $4l+2j$ final state after requesting exactly two isolated leptons,
    as outlined in
    Sec.~\ref{sec:2l}.]{\includegraphics[width=0.47\textwidth]{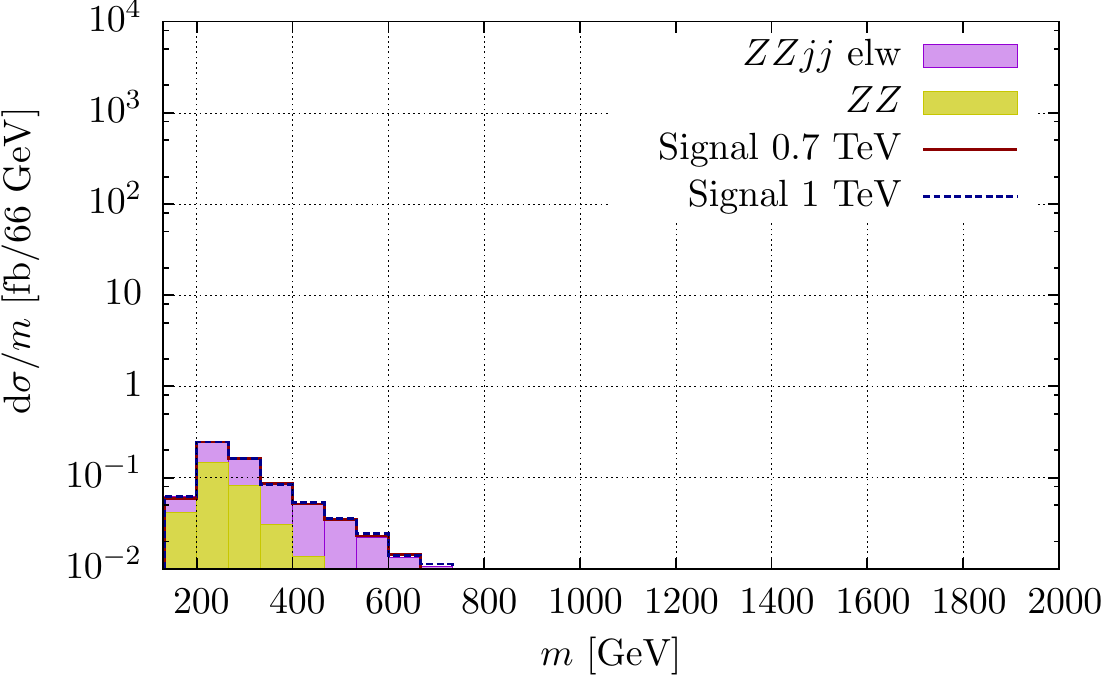}}
  \end{center}
  \caption{\label{fig:bump} Results of the WBF analysis in the $2\ell
    + \slashed{E}_T + jj$ channel (a), the $3\ell + \slashed{E}_T +
    jj$ channel (b) and the $4\ell+jj$ channel (c). All signals refer
    to a choice of $\alpha=0.9$.}
\end{figure}

\begin{figure}[!t]
  \begin{center}
        \includegraphics[width=0.47\textwidth]{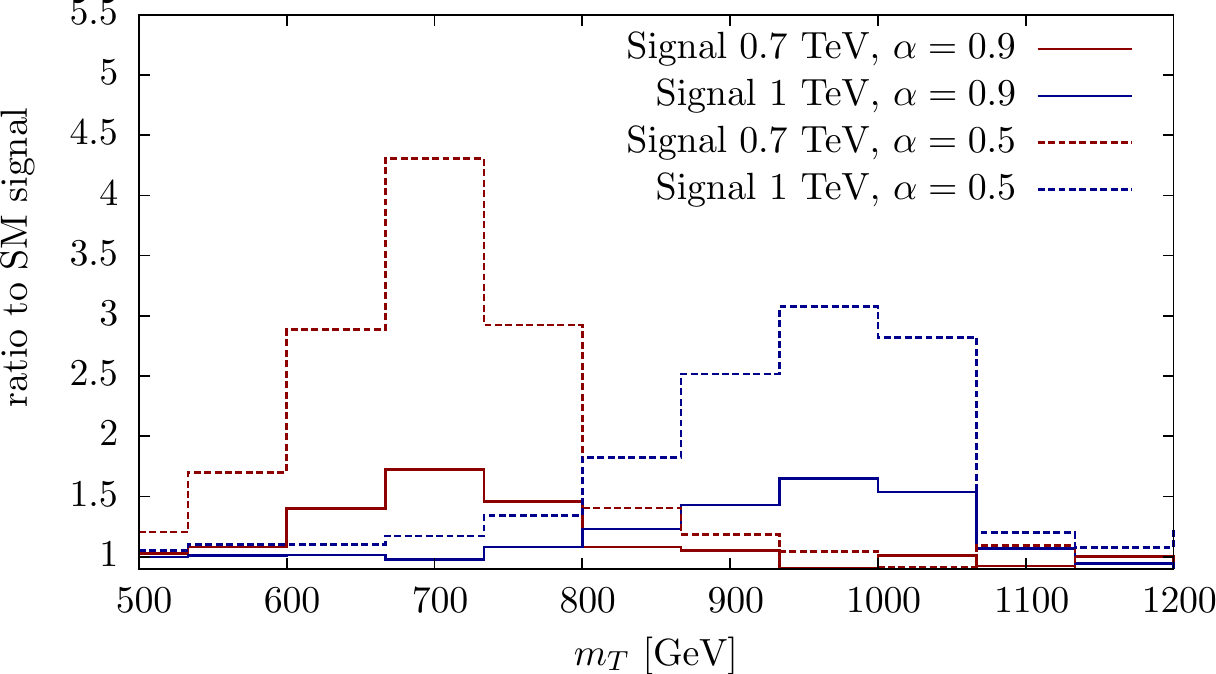}
      \caption{\label{fig:ratio}Ratio of the BSM differential cross section in $pp
          \to W^\pm Z jj \to 3\ell \slashed{E}_T jj$ in comparison
          with the SM WBF distribution. Shown are different values
          $\alpha=0.5, 0.9$; widths are chosen 3~GeV, 7 GeV, 10 GeV
          and 30 GeV, respectively.}
  \end{center}
\end{figure}

\subsection{Projections for $2l+\slashed{E}_T+jj$ production}
\label{sec:2l}
For the analysis of the $2l+\slashed{E}_T+jj$ channel, we follow the
event reconstruction outlined in Sec.~\ref{sec:sim}, and we require
exactly 2 isolated leptons. We impose staggered cuts on the transverse
momenta of both leptons, {\it i.e.}
\begin{eqnarray}
\label{eqn:lepton}
p_{T,l_1} &>& 120~\mathrm{GeV}, \nonumber\\
p_{T,l_2} &>& 80~\mathrm{GeV}. 
\end{eqnarray}

Additionally, for the two most forward jets with $p_T > 40$ GeV we
impose a WBF selection of:
\begin{eqnarray}
\label{eqn:vbf}
y_{j_1} \times y_{j_2} < 0 \nonumber \\
|y_{j_1} - y_{j_2}| > 4.0  \\
m_{j_1j_2} > 800~\mathrm{GeV} \nonumber
\end{eqnarray}

The heavy resonance is reconstructed by requiring the transverse
mass $m_T > 350$ GeV where
\begin{eqnarray}
  m_{T,2l}^2 &=& \left [ \sqrt{m^2_{l_1l_2} + p^2_{T,ll} } + |p_{T,\mathrm{miss}}| \right ]^2 \nonumber\\
  &-& \left [ {\vec{p}}_{T,ll} + {\vec{p}}_{T,\mathrm{miss}} \right ]^2.
\end{eqnarray}

We show the results after each analysis step in Tab.~\ref{tabww}. The
$WW$ channel is the most complicated final state in terms of
background composition and final state reconstruction given the
expected detector performance.

\begin{table*}[!t]
\parbox{0.55\textwidth}
{\begin{center}
\begin{tabular}{|c|c|c|c|c|}
\hline
 Sample & lepton cuts & WBF cuts & $m_{T,2l}$  \\
\hline\hline
  $(h\to WW) jj$~GF &  0.03    & $< 0.01$            &   $<0.01$             \\
 $t\bar{t}+$jets &   82.76                &   0.22           &   0.17                     \\
 $WW+$jets &  6.32                 &     1.72       &    1.09                   \\
 $WZ+$jets &   0.47                &   0.07          &  0.04                     \\
 $ZZ+$jets &     0.64              &     0.12        &   0.06                       \\
 $Z+$jets &     0.08             &      $<0.01$        &    $<0.01$                         \\
 \hline
 $m_{W',Z'}=700~\text{GeV}$, $\alpha=0.9$ &   6.37              &   1.84         &  1.24           \\
 $m_{W',Z'}=1000~\text{GeV}$, $\alpha=0.9$ &  5.89            &     1.68        &  1.18         \\ 
 $m_{W',Z'}=1500~\text{GeV}$, $\alpha=0.9$ &   5.80                &  1.67            &  1.13             \\
 $m_{W',Z'}=2000~\text{GeV}$, $\alpha=0.9$ &    5.84               &  1.64           & 1.09              \\
 $m_{W',Z'}=700~\text{GeV}$, $\alpha=0.5$  &   8.43                &    2.30          &  1.73                \\
 $m_{W',Z'}=1000~\text{GeV}$, $\alpha=0.5$ &   6.85             &   1.96           &   1.41             \\
 $m_{W',Z'}=1500~\text{GeV}$, $\alpha=0.5$ &      6.44            &   1.78           &  1.22              \\
 $m_{W',Z'}=2000~\text{GeV}$, $\alpha=0.5$ &       6.36            &   1.77         & 1.17  \\
 \hline
\end{tabular}
\end{center}}
\hskip 0.6cm
\parbox{0.35\textwidth}{
  \vskip 4.5cm
  \caption{\label{tabww}Results for 2 lepton search. The cross sections are given
    in femtobarn, corresponding to proton-proton collisions at
    $\sqrt{s} = 14$ TeV. Further details on the cuts can be found in
    the text.}
  }
\end{table*}

There are two major conclusions at this stage:

\begin{enumerate}[(i)]
\item Due to the departure of $\alpha<1$, a continuum enhancement for
  the BSM signal over the expected electroweak $VVjj$ distribution is
  present. This excess is not big enough to be useful to constrain
  this scenario efficiently; this also applies to the novel
  non-resonant $t$- and $u$-channel contributions. When we approach
  the SM limit (as supported by current measurements) the signal
  contributions quickly decouple and the analysis loses sensitivity
  even for small widths. In this sense, the phase space region
  complementary to the on-shell Higgs region cannot be efficiently
  exploited phenomenologically. Deviations from the SM WBF hypothesis
  are typically of the order of 10\%, which can easily be obstructed
  by additional experimental and theoretical systematics (see {\it
    e.g.}~\cite{width}) neglected in this analysis. The gluon fusion
  contribution is highly suppressed and we do not include it in
  Fig.~\ref{fig:bump}.
\item We therefore proceed to reconstruct the presence of $s$-channel
  resonances in a bump search sensitive to both the $WWjj$ and $ZZjj$
  subprocesses. The $2l+\slashed{E}_T+jj$ final state, however, is
  also characterized by a relatively large fraction of missing energy,
  which substantially hampers a bump search, Fig.~\ref{fig:ww}. This
  again becomes more severe when we turn to Higgs couplings in the
  vicinity of the SM expectation, see Fig.~\ref{fig:coupling}.
\end{enumerate}

\begin{table}[!t]
\begin{center}
\begin{tabular}{|c|c|c|c|c|}
\hline
 Sample & lepton cuts & WBF cuts & $m_{4l}$ \\
\hline\hline
$ZZ+$jets   &  0.25         &   0.074                         & 0.054        \\
 \hline \hline
 $\alpha=0.9$ &    0.23              &   0.075           &  0.053                  \\
 $\alpha=0.5$ &   0.24               &   0.078          &  0.058
             \\
 \hline
\end{tabular}
\end{center}
\caption{\label{tabzz}
  Results for 4 lepton search. The cross sections are given in
  femtobarn, corresponding to proton-proton collisions at $\sqrt{s} =
  14$ TeV. The $t$- and $u$-channel mass scales have no significant
  impact. Further details on the cuts can be found in the text.}
\end{table}

\subsection{Projections for WBF $4l+jj$ production}

\begin{table}[!t]
\begin{center}
\begin{tabular}{|c|c|c|c|}
\hline
 Sample & lepton cuts & WBF cuts & $m_{T,3l}$ \\
\hline\hline
 $WZ+$jets & 2.20    &    0.61 & 0.47  \\
 $t\bar{t}+$jets &   0.013   &  0         &   0                  \\
 \hline\hline
 $m_{W',Z'}=700~\text{GeV}$, $\alpha=0.9$ &    2.58               &   0.75           &  0.59                    \\
 $m_{W',Z'}=1000~\text{GeV}$, $\alpha=0.9$ &   2.32      &    0.67       &   0.51                \\
 $m_{W',Z'}=1500~\text{GeV}$, $\alpha=0.9$ &   2.22                &  0.63           &  0.48                    \\
 $m_{W',Z'}=2000~\text{GeV}$, $\alpha=0.9$ &    2.23               &  0.63            & 0.48                   \\
 $m_{W',Z'}=700~\text{GeV}$, $\alpha=0.5$ &   4.01               &   1.22          &  1.06                        \\
 $m_{W',Z'}=1000~\text{GeV}$, $\alpha=0.5$ &   2.82             &   0.84          &   0.68                  \\
 $m_{W',Z'}=1500~\text{GeV}$, $\alpha=0.5$ &      2.40             &   0.69           &  0.54                      \\
 $m_{W',Z'}=2000~\text{GeV}$, $\alpha=0.5$ &     2.31            &   0.66           &   0.50                   \\
 \hline
\end{tabular}
\end{center}
\caption{\label{tabwz}Results for 3 lepton search. The cross sections are given in femtobarn, corresponding to proton-proton collisions at $\sqrt{s} = 14$ TeV. Further details on the cuts can be found in the text.}
\end{table}

The systematic shortcomings resulting from the missing transverse
energy in WW final state are not present in the fully-reconstructible
final state $4l+jj$. We require exactly $4$ leptons and follow
Eqs.~(\ref{eqn:lepton}) and (\ref{eqn:vbf}). Additionally, the four
lepton mass is required to be $m_{4l}>350$ GeV. The cut flow is
depicted in Tab.~\ref{tabzz}. The backgrounds are manageable, however,
for the considered scenario there is no $s$-channel resonance and
again the continuum enhancement is too small to provide solid
discrimination from a non-SM realization of EWSB, if we compare the
deviations of Tab.~\ref{tabzz} to ${\cal{O}}(10\%)$ expected
experimental systematic uncertainties (see
Fig.~\ref{fig:zz}). However, this channel remains a ``golden channel''
for an additional iso-scalar resonance, and the comparison to $WW$ and
$WZ$ analyses will allow to reach a fine-grained picture of the
involved dynamics if resonances are discovered in either of the
mentioned channels.

\begin{figure*}[!t]
  \begin{center}
    \subfigure[\label{excla} 95\% confidence level (dashed) and
    $5\sigma$ discovery (solid) contours in the mass-width plane of
    the $3l+\slashed{E}_T+jj$ analysis for an integrated luminosity of
    100/fb and $\alpha=0.9$ (red) and $\alpha^2=0.95$
    (green).]{\includegraphics[width=0.47\textwidth]{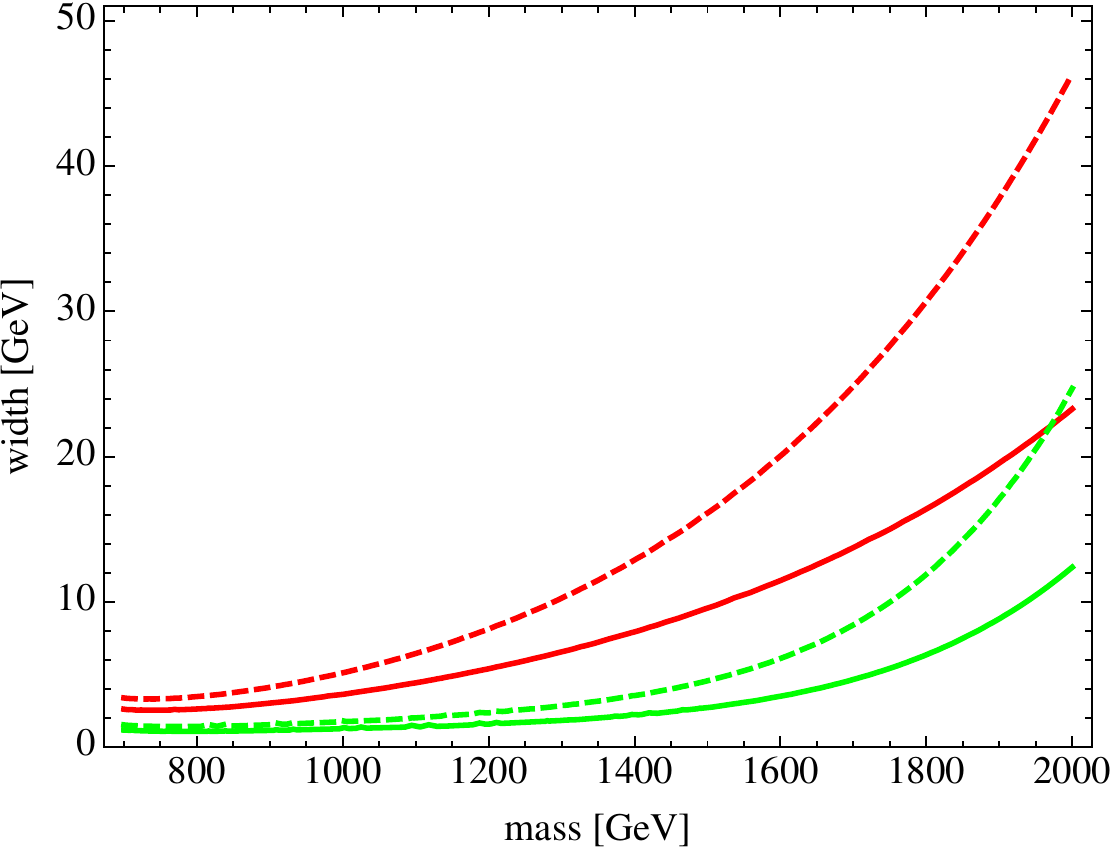}}\hfill
    \subfigure[\label{exclb} 95\% confidence level exclusion contours
    for 700 GeV (blue), 1000 GeV (red) and 1500 GeV (yellow) for a
    nominal luminosity of 100/fb.]{\includegraphics[width=0.47\textwidth]{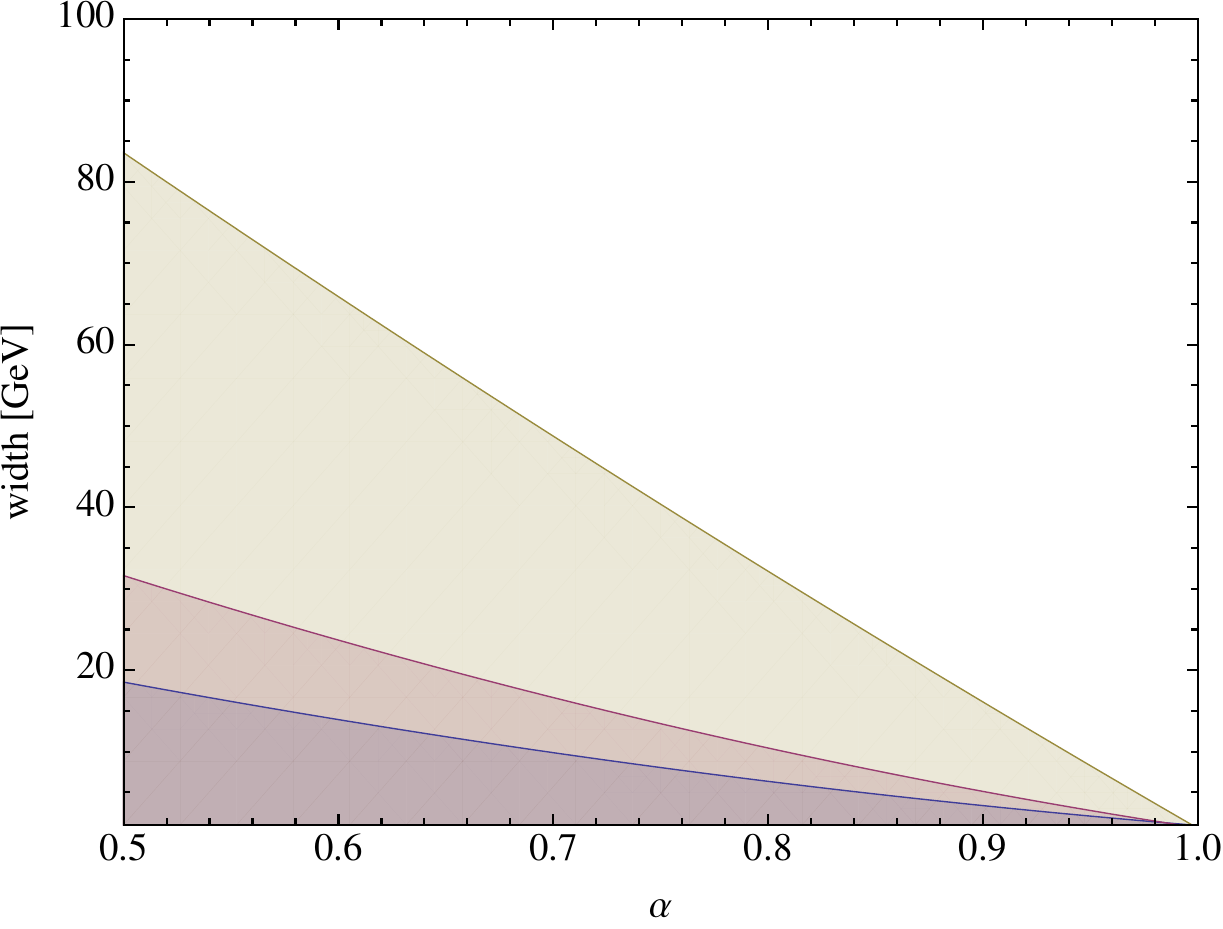}}
  \end{center}
  \vspace{-0.5cm}
  \caption{\label{fig:exl} Projections of the $3l+\slashed{E}_T+jj$
    analysis for a small integrated luminosity of 100/fb.}
\end{figure*}

\begin{figure}[!t]
  \begin{center}
    \includegraphics[width=0.47\textwidth]{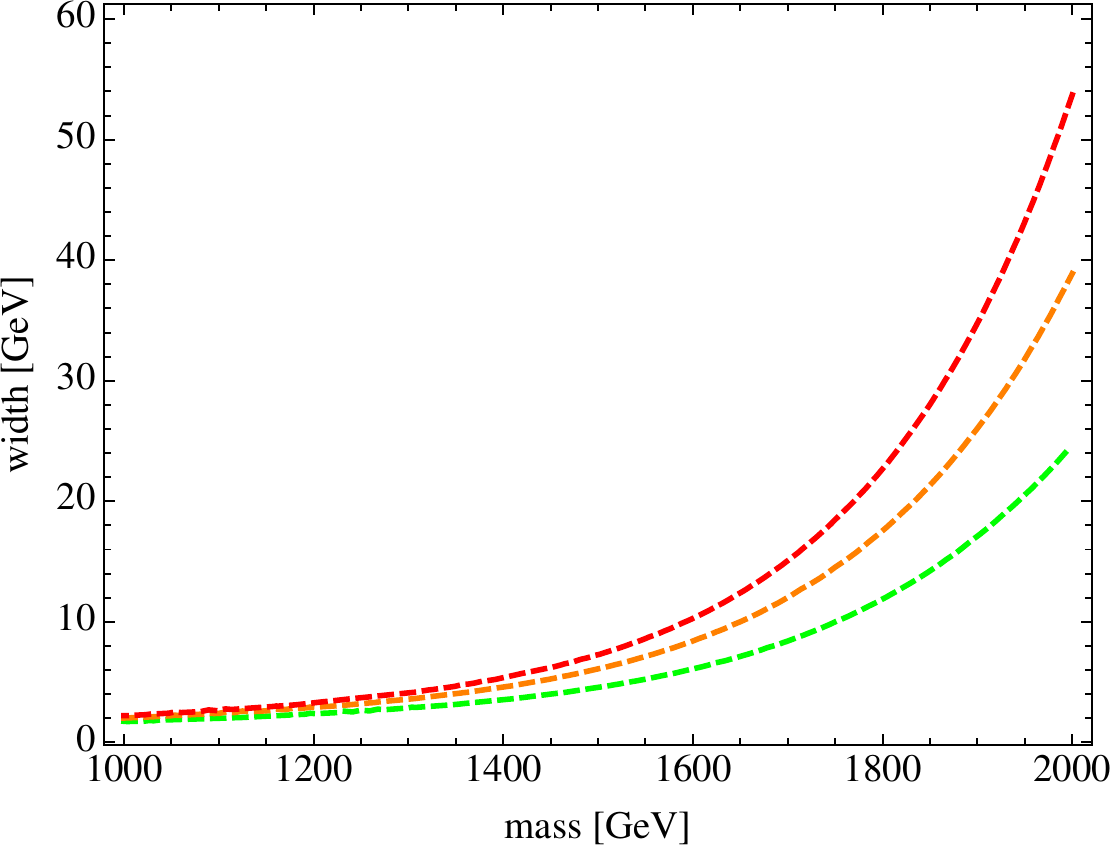}
\end{center}
\caption{\label{fig:exl2} Projections of the $3l+\slashed{E}_T+jj$
  95\% confidence level contours for 100/fb (green), 500/fb (orange)
  and 3000/fb (red). The Higgs coupling deviation is $\alpha^2=0.95$.}
\end{figure}

\subsection{Projections for $3l+\slashed{E}_T+jj$ production}
\label{sec:3l}
The $3l+\slashed{E}_T+jj$ ``interpolates'' between the previous
analyses. There is no pollution from gluon fusion events (even if we
allow a significant coupling of $Z'$ to the fermion
sector). Additionally, the major backgrounds of the
$2l+\slashed{E}_T+jj$ can be completely removed through the
requirement of exactly three isolated leptons with $p_{T,l} > 15$ GeV,
with no charge requirement. We then require the cuts given in
Eqs.~(\ref{eqn:lepton}) and (\ref{eqn:vbf}) and for the lepton and WBF
selection respectively.  The signal is extracted following a final
selection $m_{T,3l} > 350$ GeV, where
\begin{eqnarray}
  \label{eq:mtwz}
  m^2_{T,3l}&=& \left [ \sqrt{m^2_{l_1l_2l_3} + p^2_{T,l_1l_2l_3} } 
    + |p_{T,\mathrm{miss}}| \right ]^2 \nonumber\\
  &-& \left [ {\vec{p}}_{T,l_1l_2l_3} + {\vec{p}}_{T,\mathrm{miss}} \right ]^2.
\end{eqnarray}
The results are collected in Tab.~\ref{tabwz}.

Although a substantial amount of missing energy is present, the
lepton-$\slashed{E}_T$ system is highly correlated in this final
state, allowing for recovery of most of the mass discrimination
through Eq.~\eqref{eq:mtwz}, see Fig.~\ref{fig:wz}.

As a result of the non-standard Higgs coupling, a large enhancement of
the signal strength is present. This can be seen compared to the
standard model background in Fig.~\ref{fig:ratio}.

\subsection{Setting limits with $3l+\slashed{E}_T+jj$ production}
\label{sec:limits}

Combining the analyses of the previous sections, we can see that the
potential presence of new vector resonances for $\sim 10\%$ Higgs
coupling deviations can be highly constrained with the
$3l+\slashed{E}_T+jj$ channel. Although we believe that more advanced
limit setting procedures that deal with full correlations can
eventually be used to constrain iso-triplet states in the
$2l+{\slashed{E}}_T+jj$ and $4l+jj$ final states, the
$3l+\slashed{E}_T+jj$ provides the most direct avenue to constrain
such a scenario.

We thus quote an expected significance using
  $3l+\slashed{E}_T+jj$ final states (Sec.~\ref{sec:3l}) on the basis
  of mass, width and modified Higgs coupling strength in
  Figs.~\ref{excla} and \ref{exclb}. The signal extraction is
performed over a mass window of $0.3\times m_{W'}$ in the transverse
mass Eq.~\eqref{eq:mtwz}. The calculated significance follows from:
\begin{equation}
  {\cal{S}}= {N(\text{BSM})-N({\text{WBF,SM}})\over \sqrt{N(\text{bkg,non-WBF})+N({\text{WBF,SM}})}}\,,
\end{equation}
where the individual $N$s refer to the signal counts at a given
luminosity. Using this measure we can isolate a statistically
significant deviation from the SM WBF distribution outside the Higgs
signal region, taking into account the irreducible background in the
$WZ$ channel.

Already for a target luminosity of run 2 of 100/fb, a large parameter
region can be explored in the $3l+\slashed{E}_T+jj$ channel. A crucial
parameter in this analysis is the width of the additional resonance,
which we take as a free parameter in our analysis. With an increasing
width the signal decouples quickly, but stringent constraints can
still be formulated at a high-luminosity LHC, especially if new
physics gives rise to only a percent-level deformation of the SM Higgs
interactions, see Fig.~\ref{fig:exl2}. Note that the signal decouples
very quickly with an increased value of the width. Hence, if there in
scenarios where the extra vector bosons have a large coupling to the
top as expected in some composite models, the sensitivity in the WBF
search might not be sufficient to constrain the presence of such
states. It is worthwhile to stress the complementarity of the WBF
searches as outlined in the previous sections to the aforementioned
Drell-Yan like production in this regard. Both ATLAS and CMS have
published limits of searches for $W'$ and $Z'$ resonances in third
quark generation final states \cite{atlasdy,cmsdy,atlasdy2,cmsdy2}. If
the states we investigate in this paper have a sizeable coupling to
massive fermions, these searches will eventually facilitate a
discovery. In this case, however, the search for WBF resonances still
provides complementary information about the nature of electroweak
symmetry breaking. In particular WBF production will act as a
consistency check of the excesses around 2 TeV seen by CMS and
ATLAS~\cite{atlasex,cmsex}. 

\begin{figure}[!b]
  \begin{center}
    \includegraphics[width=0.47\textwidth]{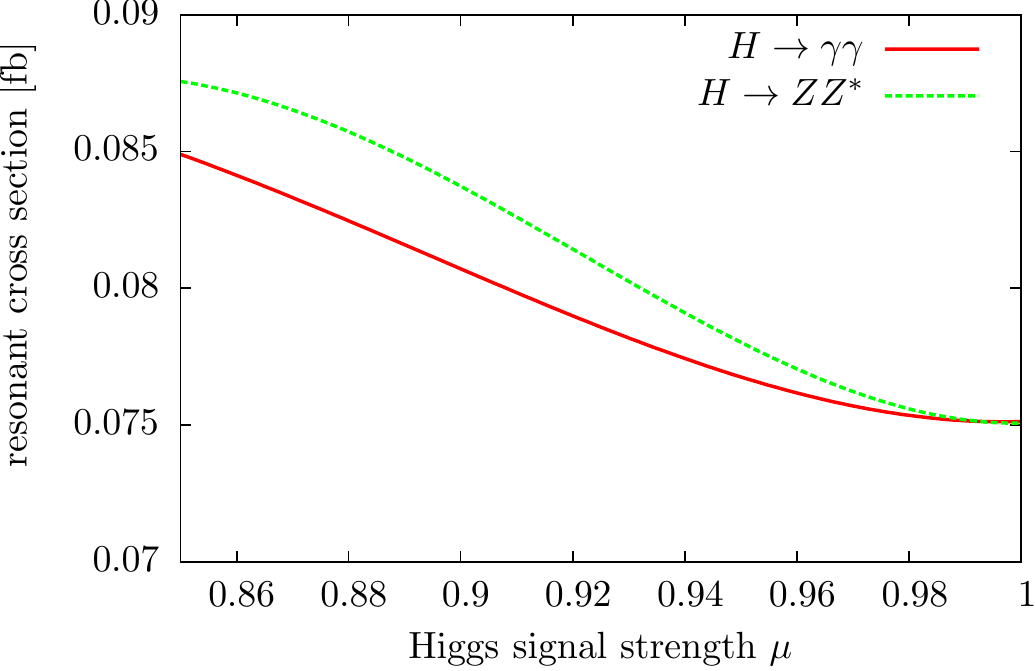}
\end{center}
\caption{\label{fig:am1} Cross section of 2 TeV diboson resonance in
  WBF for single lepton inclusive cuts at 8 TeV center of mass energy.}
\end{figure}

In Fig.~\ref{fig:am1} we show the cross section for a 2 TeV resonance
in WBF correlated with the Higgs boson on-shell signal strengths for
the scenario where the extra resonances width solely arises from the
partial width to SM gauge bosons. This is optimistic in the sense that
the expected signal rate is maximised; the Higgs phenomenology is only
modified via the interactions with the gauge bosons (see above). As
can be seen from the inclusive cross section in Fig.~\ref{fig:am1} the
expected cross section before reconstruction is far to small to
account for a $\sim 1$~fb signal cross section required to explain the
ATLAS and CMS anomalies. If these excesses become statistically
significant, this means that the observed particle(s) do not stand in
relation relation to longitudinal gauge boson
unitarization. Alternative scenarios are discussed
in~\cite{Cacciapaglia:2015eea}.

\section{Summary and Conclusions}
\label{sec:conc}
The search for new physics interactions after the discovery of the
Higgs boson remains one of the main targets of the LHC. Current
constraints on Higgs couplings inferred from run 1 signal strength
measurements, in particular in the $ZZ$ channel, leave a lot of space
for the appearance of new resonant phenomena at the TeV scale. These
can, but not necessarily have to be iso-scalar degrees of freedom. To
this end we have combined the observation of a SM-like Higgs boson
with the appearance of new iso-vectorial degrees of freedom at the TeV
scale. These are further corroborated by small excesses in similar and
recent searches during run 1~\cite{exibump}. Solely based on
probability conservation, we provide predictions for the weak boson
fusion channels, which are theoretically well motivated candidate
processes to study resonant phenomena connected to unitarity and the
anatomy of electroweak symmetry breaking. Our approach of saturating
$W,Z$ unitarity sum rules with a single set of vector resonances as a
function of vector boson mass and Higgs coupling deviation provides a
complementary approach to singlet-extended Higgs sectors with a highly
modified TeV-scale LHC phenomenology.

While resonances and continuum excesses due to new $t$- and $u-$
channel contributions and a smaller destructive Higgs contribution at
large multi-lepton mass might be challenging to observe in
$2l+{\slashed{E}}_T+jj$ and $4l+jj$ production, we have shown that an
analysis of $3l+{\slashed{E}}_T+jj$ production provides an excellent
avenue to constrain or even observe the presence of such states over a
broad range of mass and width scales. With comparably low integrated
luminosity at the LHC, such an analysis captures complimentary and
necessary information to pin down the very character of new physics
for small deviations of the Higgs on-shell phenomenology, especially
when results across the different WBF channels are combined.

\bigskip 

{\bf{Acknowledgments}} --- CE is supported by the Institute for
Particle Physics Phenomenology Associate scheme. We thank Sven
Heinemeyer for helpful conversations.  This research was supported in
part by the European Commission through the ``HiggsTools'' Initial
Training Network PITN-GA-2012-316704. This work was supported by World
Premier International Research Center Initiative (WPI Initiative),
MEXT, Japan.



\begin{thebibliography}{99}

\bibitem{orig} 
  F.~Englert and R.~Brout,
  Phys.\ Rev.\ Lett.\  {\bf 13} (1964) 321.
  P.~W.~Higgs,
  Phys.\ Lett.\  {\bf 12} (1964) 132 and
  Phys.\ Rev.\ Lett.\  {\bf 13} (1964) 508.
  G.~S.~Guralnik, C.~R.~Hagen and T.~W.~B.~Kibble,
  Phys.\ Rev.\ Lett.\  {\bf 13} (1964) 585.

\bibitem{hatlas}
  G.~Aad {\it et al.}  [ATLAS Collaboration],
  Phys.\ Lett.\ B {\bf 716} (2012) 1
  [arXiv:1207.7214 [hep-ex]].

\bibitem{hcms}
  S.~Chatrchyan {\it et al.}  [CMS Collaboration],
  Phys.\ Lett.\ B {\bf 716} (2012) 30
  [arXiv:1207.7235 [hep-ex]].


\bibitem{sven}
  M.~Duhrssen, S.~Heinemeyer, H.~Logan, D.~Rainwater, G.~Weiglein and D.~Zeppenfeld,
  Phys.\ Rev.\ D {\bf 70} (2004) 113009
  [hep-ph/0406323].
  B.~A.~Dobrescu and J.~D.~Lykken,
  JHEP {\bf 1302} (2013) 073
  [arXiv:1210.3342 [hep-ph]].
  P.~Bechtle, S.~Heinemeyer, O.~Stal, T.~Stefaniak and G.~Weiglein,
  JHEP {\bf 1411} (2014) 039
  [arXiv:1403.1582 [hep-ph]].
  J.~Ellis, V.~Sanz and T.~You,
  JHEP {\bf 1407} (2014) 036
  [arXiv:1404.3667 [hep-ph]].
  J.~Ellis, V.~Sanz and T.~You,
  arXiv:1410.7703 [hep-ph].

\bibitem{sfitter}
  D.~Lopez-Val, T.~Plehn and M.~Rauch,
  JHEP {\bf 1310} (2013) 134
  [arXiv:1308.1979 [hep-ph]].
  C.~Englert, A.~Freitas, M.~M.~Mühlleitner, T.~Plehn, M.~Rauch, M.~Spira and K.~Walz,
  J.\ Phys.\ G {\bf 41} (2014) 113001
  [arXiv:1403.7191 [hep-ph]].

\bibitem{Giudice:2007fh}
  G.~F.~Giudice, C.~Grojean, A.~Pomarol and R.~Rattazzi,
  JHEP {\bf 0706} (2007) 045
  [hep-ph/0703164].

\bibitem{conformal}
  C.~Englert, J.~Jaeckel, V.~V.~Khoze and M.~Spannowsky,
  JHEP {\bf 1304} (2013) 060
  [arXiv:1301.4224 [hep-ph]].
  M.~Heikinheimo, A.~Racioppi, M.~Raidal and C.~Spethmann,
  Phys.\ Lett.\ B {\bf 726} (2013) 781
  [arXiv:1307.7146].
  J.~D.~Clarke, R.~Foot and R.~R.~Volkas,
  JHEP {\bf 1402} (2014) 123
  [arXiv:1310.8042 [hep-ph]].
  A.~Farzinnia and J.~Ren,
  arXiv:1405.0498 [hep-ph].

\bibitem{Bellazzini:2012tv}
  B.~Bellazzini, C.~Csaki, J.~Hubisz, J.~Serra and J.~Terning,
  JHEP {\bf 1211} (2012) 003
  [arXiv:1205.4032 [hep-ph]].

\bibitem{gaugephobic}
  G.~Cacciapaglia, C.~Csaki, G.~Marandella and J.~Terning,
  JHEP {\bf 0702} (2007) 036
  [hep-ph/0611358].
  J.~Galloway, B.~McElrath, J.~McRaven and J.~Terning,
  JHEP {\bf 0911} (2009) 031
  [arXiv:0908.0532 [hep-ph]].


\bibitem{Cornwall:1974km}
  J.~M.~Cornwall, D.~N.~Levin and G.~Tiktopoulos,
  Phys.\ Rev.\ Lett.\  {\bf 30} (1973) 1268
  [Erratum-ibid.\  {\bf 31} (1973) 572].
  J.~M.~Cornwall, D.~N.~Levin and G.~Tiktopoulos,
  Phys.\ Rev.\ D {\bf 10} (1974) 1145
  [Erratum-ibid.\ D {\bf 11} (1975) 972].

\bibitem{wbf_old}
  V.~D.~Barger, K.~-m.~Cheung, T.~Han and D.~Zeppenfeld,
  Phys.\ Rev.\ D {\bf 44} (1991) 2701
   [Erratum-ibid.\ D {\bf 48} (1993) 5444].
  J.~Bagger, V.~D.~Barger, K.~-m.~Cheung, J.~F.~Gunion, T.~Han, G.~A.~Ladinsky, R.~Rosenfeld and C.~-P.~Yuan,
  Phys.\ Rev.\ D {\bf 52} (1995) 3878
  [hep-ph/9504426].
  D.~L.~Rainwater and D.~Zeppenfeld,
  Phys.\ Rev.\ D {\bf 60} (1999) 113004
  [Erratum-ibid.\ D {\bf 61} (2000) 099901]
  [hep-ph/9906218].
  N.~Kauer, T.~Plehn, D.~L.~Rainwater and D.~Zeppenfeld,
  Phys.\ Lett.\ B {\bf 503} (2001) 113
  [hep-ph/0012351].
  C.~Englert, B.~Jager, M.~Worek and D.~Zeppenfeld,
  Phys.\ Rev.\ D {\bf 80} (2009) 035027
  [arXiv:0810.4861 [hep-ph]].

\bibitem{wbf_new}
  A.~Ballestrero, D.~B.~Franzosi, L.~Oggero and E.~Maina,
  JHEP {\bf 1203} (2012) 031
  [arXiv:1112.1171 [hep-ph]].
  P.~Borel, R.~Franceschini, R.~Rattazzi and A.~Wulzer,
  JHEP {\bf 1206} (2012) 122
  [arXiv:1202.1904 [hep-ph]].
  A.~Freitas and J.~S.~Gainer,
  Phys.\ Rev.\ D {\bf 88} (2013) 1,  017302
  [arXiv:1212.3598].
  R.~L.~Delgado, A.~Dobado and F.~J.~Llanes-Estrada,
  Phys.\ Rev.\ D {\bf 91} (2015) 7,  075017
  [arXiv:1502.04841 [hep-ph]].

\bibitem{Englert:2014aca} 
  C.~Englert and M.~Spannowsky,
  Phys.\ Rev.\ D {\bf 90}, no. 5, 053003 (2014)
  [arXiv:1405.0285 [hep-ph]];
  A.~Biekoetter, A.~Knochel, M.~Kraemer, D.~Liu and F.~Riva,
  arXiv:1406.7320 [hep-ph].


\bibitem{dim6}
  B.~Grzadkowski, M.~Iskrzynski, M.~Misiak and J.~Rosiek,
  JHEP {\bf 1010} (2010) 085
  [arXiv:1008.4884 [hep-ph]].

\bibitem{perelstein_matchev}
  A.~Birkedal, K.~Matchev and M.~Perelstein,
  Phys.\ Rev.\ Lett.\  {\bf 94} (2005) 191803
  [hep-ph/0412278].

\bibitem{csaki}
  C.~Csaki, C.~Grojean, H.~Murayama, L.~Pilo and J.~Terning,
  Phys.\ Rev.\ D {\bf 69} (2004) 055006
  [hep-ph/0305237].
  C.~Csaki, C.~Grojean, L.~Pilo and J.~Terning,
  Phys.\ Rev.\ Lett.\  {\bf 92} (2004) 101802
  [hep-ph/0308038].
  C.~Csaki, J.~Hubisz and P.~Meade,
  hep-ph/0510275.

\bibitem{Bhattacharyya:2012tj}
  G.~Bhattacharyya, D.~Das and P.~B.~Pal,
  Phys.\ Rev.\ D {\bf 87} (2013) 1,  011702
  [arXiv:1212.4651 [hep-ph]].



\bibitem{juergen}
  A.~Alboteanu, W.~Kilian and J.~Reuter,
  JHEP {\bf 0811} (2008) 010
  [arXiv:0806.4145 [hep-ph]].


\bibitem{exibump}
  S.~Chatrchyan {\it et al.}  [CMS Collaboration],
  Phys.\ Rev.\ D {\bf 90}, no. 3, 032006 (2014)
  [arXiv:1404.5801 [hep-ex]].
  G.~Aad {\it et al.}  [ATLAS Collaboration],
  arXiv:1503.03290 [hep-ex].
  S.~Chatrchyan {\it et al.}  [CMS Collaboration], CMS PAS HIG-14-008.
  S.~Chatrchyan {\it et al.}  [CMS Collaboration], CMS EXO-12-041.


\bibitem{atlasex}
  G.~Aad {\it et al.}  [ATLAS Collaboration],
  Eur.\ Phys.\ J.\ C {\bf 75} (2015) 2,  69
  [arXiv:1409.6190 [hep-ex]].
  G.~Aad {\it et al.}  [ATLAS Collaboration],
  Eur.\ Phys.\ J.\ C {\bf 75} (2015) 5,  209
  [arXiv:1503.04677 [hep-ex]].
  G.~Aad {\it et al.}  [ATLAS Collaboration],
  arXiv:1506.00962 [hep-ex].

\bibitem{cmsex}
The CMS collaboration, EXO-13-009 and EXO-12-024.

\bibitem{Butterworth:2008iy} 
  J.~M.~Butterworth, A.~R.~Davison, M.~Rubin and G.~P.~Salam,
  Phys.\ Rev.\ Lett.\  {\bf 100}, 242001 (2008)
  [arXiv:0802.2470 [hep-ph]].


\bibitem{portal}
  T.~Binoth and J.~J.~van der Bij,
  Z.\ Phys.\ C {\bf 75} (1997) 17
  [hep-ph/9608245].
  M.~Bowen, Y.~Cui and J.~D.~Wells,
  JHEP {\bf 0703} (2007) 036
  [hep-ph/0701035].
  C.~Englert, T.~Plehn, D.~Zerwas and P.~M.~Zerwas,
  Phys.\ Lett.\ B {\bf 703} (2011) 298
  [arXiv:1106.3097 [hep-ph]].
  E.~Weihs and J.~Zurita,
  JHEP {\bf 1202} (2012) 041
  [arXiv:1110.5909 [hep-ph]].

\bibitem{trip}
  S.~Godfrey and K.~Moats,
  Phys.\ Rev.\ D {\bf 81} (2010) 075026
  [arXiv:1003.3033 [hep-ph]].
  R.~Killick, K.~Kumar and H.~E.~Logan,
  Phys.\ Rev.\ D {\bf 88} (2013) 033015
  [arXiv:1305.7236 [hep-ph]].
  C.~Englert, E.~Re and M.~Spannowsky,
  Phys.\ Rev.\ D {\bf 88} (2013) 035024
  [arXiv:1306.6228 [hep-ph]].
  C.~W.~Chiang, A.~L.~Kuo and K.~Yagyu,
  JHEP {\bf 1310} (2013) 072
  [arXiv:1307.7526 [hep-ph]].

\bibitem{chanowitz}
  M.~S.~Chanowitz, M.~A.~Furman and I.~Hinchliffe,
  Phys.\ Lett.\ B {\bf 78} (1978) 285.
  M.~S.~Chanowitz, M.~A.~Furman and I.~Hinchliffe,
  Nucl.\ Phys.\ B {\bf 153} (1979) 402.

\bibitem{drell}
H.~-J.~He, Y.~-P.~Kuang, Y.~-H.~Qi, B.~Zhang, A.~Belyaev, R.~S.~Chivukula, N.~D.~Christensen and A.~Pukhov {\it et al.},
  Phys.\ Rev.\ D {\bf 78} (2008) 031701
  [arXiv:0708.2588 [hep-ph]].
T.~Ohl and C.~Speckner,
  Phys.\ Rev.\ D {\bf 78} (2008) 095008
  [arXiv:0809.0023 [hep-ph]].

\bibitem{ricardo}
  D.~Pappadopulo, A.~Thamm, R.~Torre and A.~Wulzer,
  JHEP {\bf 1409} (2014) 060
  [arXiv:1402.4431 [hep-ph]].

\bibitem{dieter}
  V.~D.~Barger, R.~J.~N.~Phillips and D.~Zeppenfeld,
  Phys.\ Lett.\ B {\bf 346} (1995) 106
  [hep-ph/9412276].
  J.~R.~Andersen, K.~Arnold and D.~Zeppenfeld,
  JHEP {\bf 1006} (2010) 091
  [arXiv:1001.3822 [hep-ph]].
  J.~R.~Andersen, C.~Englert and M.~Spannowsky,
  Phys.\ Rev.\ D {\bf 87} (2013) 1,  015019
  [arXiv:1211.3011 [hep-ph]].

\bibitem{quartic}
P.~Achard {\it et al.}  [L3 Collaboration],
  Phys.\ Lett.\ B {\bf 527} (2002) 29
  [hep-ex/0111029].
G.~Abbiendi {\it et al.}  [OPAL Collaboration],
  Phys.\ Lett.\ B {\bf 580} (2004) 17
  [hep-ex/0309013].
J.~Abdallah {\it et al.}  [DELPHI Collaboration],
  Eur.\ Phys.\ J.\ C {\bf 31} (2003) 139
  [hep-ex/0311004].
  G.~Abbiendi {\it et al.}  [OPAL Collaboration],
  Phys.\ Rev.\ D {\bf 70} (2004) 032005
  [hep-ex/0402021].


\bibitem{vbfnlo}
  K.~Arnold, M.~Bahr, G.~Bozzi, F.~Campanario, C.~Englert, T.~Figy, N.~Greiner and C.~Hackstein {\it et al.},
  Comput.\ Phys.\ Commun.\  {\bf 180} (2009) 1661
  [arXiv:0811.4559 [hep-ph]].

\bibitem{wbfscales}
  B.~Jager, C.~Oleari and D.~Zeppenfeld,
  JHEP {\bf 0607} (2006) 015
  [hep-ph/0603177].
  G.~Bozzi, B.~Jager, C.~Oleari and D.~Zeppenfeld,
  Phys.\ Rev.\ D {\bf 75} (2007) 073004
  [hep-ph/0701105].

\bibitem{Englert:2008wp}
  C.~Englert, B.~Jager and D.~Zeppenfeld,
  JHEP {\bf 0903} (2009) 060
  [arXiv:0812.2564 [hep-ph]].

\bibitem{Stuart:1991xk}
  S.~Willenbrock and G.~Valencia,
  Phys.\ Lett.\ B {\bf 259} (1991) 373;
  R.~G.~Stuart,
  Phys.\ Lett.\ B {\bf 262} (1991) 113;
  U.~Baur and D.~Zeppenfeld,
  Phys.\ Rev.\ Lett.\  {\bf 75} (1995) 1002.
  J.~Papavassiliou and A.~Pilaftsis,
  Phys.\ Rev.\ Lett.\  {\bf 80} (1998) 2785
  [hep-ph/9710380];
  J.~Papavassiliou and A.~Pilaftsis,
  Phys.\ Rev.\ D {\bf 54} (1996) 5315
  [hep-ph/9605385].
  Y.~Bai and W.~Y.~Keung,
  arXiv:1407.6355 [hep-ph].



\bibitem{Bahr:2008pv}
  M.~Bahr, S.~Gieseke, M.~A.~Gigg, D.~Grellscheid, K.~Hamilton, O.~Latunde-Dada, S.~Platzer and P.~Richardson {\it et al.},
  Eur.\ Phys.\ J.\ C {\bf 58} (2008) 639
  [arXiv:0803.0883 [hep-ph]].

\bibitem{alpgen}
  M.~L.~Mangano, M.~Moretti, F.~Piccinini, R.~Pittau and A.~D.~Polosa,
  JHEP {\bf 0307} (2003) 001
  [hep-ph/0206293].

\bibitem{detector}
  The ATLAS collaboration, ATL-PHYS-PUB-2013-004.


\bibitem{Cacciari:2008gp} 
  M.~Cacciari, G.~P.~Salam and G.~Soyez,
  JHEP {\bf 0804}, 063 (2008)
  [arXiv:0802.1189 [hep-ph]].

\bibitem{width}
  S.~Willenbrock and G.~Valencia,
  Phys.\ Lett.\ B {\bf 259} (1991) 373.
  R.~G.~Stuart,
  Phys.\ Lett.\ B {\bf 262} (1991) 113.
  M.~Nowakowski and A.~Pilaftsis,
  Z.\ Phys.\ C {\bf 60} (1993) 121
  [hep-ph/9305321].
  U.~Baur and D.~Zeppenfeld,
  Phys.\ Rev.\ Lett.\  {\bf 75} (1995) 1002
  [hep-ph/9503344].


\bibitem{atlasdy}
  G.~Aad {\it et al.}  [ATLAS Collaboration],
  Eur.\ Phys.\ J.\ C {\bf 75} (2015) 4,  165
  [arXiv:1408.0886 [hep-ex]].
  G.~Aad {\it et al.}  [ATLAS Collaboration],
  Phys.\ Lett.\ B {\bf 743} (2015) 235
  [arXiv:1410.4103 [hep-ex]].

\bibitem{cmsdy}
  S.~Chatrchyan {\it et al.}  [CMS Collaboration],
  JHEP {\bf 1405} (2014) 108
  [arXiv:1402.2176 [hep-ex]].
  V.~Khachatryan {\it et al.}  [CMS Collaboration],
  Phys.\ Rev.\ D {\bf 91} (2015) 9,  092005
  [arXiv:1408.2745 [hep-ex]].


\bibitem{atlasdy2}
  The ATLAS Collaboration, ATLAS-CONF-2015-009

\bibitem{cmsdy2}
  The CMS Collaboration, CMS-PAS-B2G-12-007.


\bibitem{Cacciapaglia:2015eea}
  D.~B.~Franzosi, M.~T.~Frandsen and F.~Sannino,
  arXiv:1506.04392 [hep-ph].
  J.~A.~Aguilar-Saavedra,
  arXiv:1506.06739 [hep-ph].
  J.~Hisano, N.~Nagata and Y.~Omura,
  arXiv:1506.03931 [hep-ph].
  K.~Cheung, W.~Y.~Keung, P.~Y.~Tseng and T.~C.~Yuan,
  arXiv:1506.06064 [hep-ph].
  B.~A.~Dobrescu and Z.~Liu,
  arXiv:1506.06736 [hep-ph].
  T.~Abe, T.~Kitahara and M.~M.~Nojiri,
  arXiv:1507.01681 [hep-ph].
  J.~Brehmer, J.~Hewett, J.~Kopp, T.~Rizzo and J.~Tattersall,
  arXiv:1507.00013 [hep-ph].
  G.~Cacciapaglia and M.~T.~Frandsen,
  arXiv:1507.00900 [hep-ph].
  B.~C.~Allanach, B.~Gripaios and D.~Sutherland,
  arXiv:1507.01638 [hep-ph].
  A.~Carmona, A.~Delgado, M.~Quiros and J.~Santiago,
  arXiv:1507.01914 [hep-ph].

\end{thebibliography}
\end{document}